\documentclass[a4paper,12pt]{article}
\usepackage{amsmath,amssymb,graphicx}
\newcommand\figcaption{\def\@captype{figure}\caption}
\newcommand\tabcaption{\def\@captype{table}\caption}
\def\pa{\partial}
\def\al{\alpha}
\def\ga{\gamma}
\def\Ga{\Gamma}

\def\Dl{\Delta}

\def\be{\beta}
\def\kp{\kappa}
\def\th{\theta}
\def\sg{\sigma}
\def\nb{\nabla}
\def\vf{\varphi}
\def\ck{\check}
\def\diag{\mbox {diag}}
\def\nn{\nonumber}

\begin{document}
\title{\bf Eigen Equation of the Nonlinear Spinor}
\author{Ying-Qiu Gu\footnote{email: yqgu@luody.com.cn} \qquad Ta-tsien Li\footnote {email: dqli@fudan.edu.cn}}
\date{\small Department of Mathematics, Fudan University,
Shanghai, 200433, China}

\maketitle\DeclareGraphicsRule{.eps.gz}{eps}{.eps.bb}{`gunzip -c
#1}

\begin{abstract}
How to effectively solve the eigen solutions of the nonlinear
spinor field equation coupling with some other interaction fields
is important to understand the behavior of the elementary
particles. In this paper, we derive a simplified form of the eigen
equation of the nonlinear spinor, and then propose a scheme to
solve their numerical solutions. This simplified equation has
elegant and neat structure, which is more convenient for both
theoretical analysis and numerical computation.

\vskip 1.0cm \large {PACS numbers: 11.10.Ef, 11.10.Lm, 11.10.-z}

\vskip 1.0cm \large{Key Words: {\sl nonlinear Dirac equation,
nonlinear spinor field, eigen equation, quaternion}}
\end{abstract}

\section{Introduction}
\setcounter{equation}{0}

Almost all the elementary fermions have spin-$\frac 1 2$, which
can be naturally described by spinors, so today spinors and spinor
representations play a more and more important role in
mathematical and theoretical physics. Noticing the limitations of
the linear field equation, many physicists such as H. Weyl, W.
Heisenberg, once proposed the nonlinear spinor
equations\cite{1,2,3,4,5,6} to construct a unified field theory
for elementary particles. However they have not gotten many
definite results due to the mathematical difficulties. The
rigorous solutions for some simple dark nonlinear spinor models
were obtained in \cite{7,8,gu1,gu2}, and we found it can provide
negative pressure to guarantee a singularity-free and accelerating
expanding universe\cite{gu3,gu4}.

The theoretical proof about the existence of solitons was
investigated in \cite{12,13,14,15,16,17}. The symmetries and many
beautiful conditional exact solutions of the nonlinear spinor,
vector and scalar differential equations are collected in
\cite{18}. However lots of these exact solutions seem to be
non-physical.

The spinor with its own electromagnetic potential was studied in
\cite{19,20,gu5,bar1,bar2}, it was disclosed that the nonlinear
spinor equations have particle like solution with anomalous
magneton, and imply the exact classical mechanics and quantum
mechanics for many-body\cite{gu6,gu7,gu8}.

In this paper we derive a simplified form of the eigen equation
with general meaning for nonlinear Dirac equation, and then give a
scheme to solve the solution. Denote the Minkowski metric by
$\eta_{\mu\nu}={\rm diag}[1,-1,-1,-1]$, Pauli matrices by
\begin{equation}
 {\vec\sg}=(\sg^{j})= \left\{\left(\begin{matrix}
 0 & 1 \cr 1 & 0\end{matrix}\right),~\left(\begin{matrix}
 0 & -i \cr i & 0\end{matrix}\right),~\left(\begin{matrix}
 1 & 0 \cr 0 & -1\end{matrix}\right)\right\}
 .\label{1.1}\end{equation}
Define $4\times4$ Hermitian matrices as follows
\begin{equation}\al^\mu=\left\{\left ( \begin{array}{ll} I & ~0 \\
0 & I \end{array} \right),\left (\begin{array}{ll} 0 & \vec\sg \\
\vec\sg & 0 \end{array}
\right)\right\},~ \ga =\left ( \begin{array}{ll} I & ~0 \\
0 & -I \end{array} \right),
~ \be=\left (\begin{array}{ll} 0 & -iI \\
iI & ~~0 \end{array} \right).\label{1.2}
\end{equation}
In this paper, we adopt the Hermitian matrices (\ref{1.2}) instead
of Dirac matrices $\ga^\mu$, because this form is more convenient
for calculation.

For the system of a nonlinear spinor field $\phi$ in the $4-d$
potential $A^{\mu}$, the Lagrangian describing the motion is
generally given by
\begin{equation}
{\cal L} =\phi^+[\al^\mu (\hbar i\pa_\mu-eA_\mu) -\mu c
\ga]\phi+F(\ck\ga,\ck\be), \label{1.3}\end{equation} where $\mu>0$
is a constant mass, $F$ are the nonlinear coupling potential,
which is usually the even polynomial of $\ck\ga,\ck\be$, and
$\ck\ga,\ck\be$ are the quadratic scalars of $\phi$ defined by
\begin{equation}
\ck\ga=\phi^{+}\ga\phi, \qquad \ck\be=\phi^{+}\be\phi.
\label{1.4}\end{equation} We can check $\ck\ga$ is a true-scalar,
but $\ck\be$ a pseudo-scalar.

The variation of (\ref{1.3}) with respect to $\phi^+$ gives the
dynamic equation
\begin{equation}
\al^\mu(\hbar i\pa_\mu-eA_\mu)\phi=(\mu c \ga -F_\ga\ga-F_\be\be)
\phi, \label{1.5}
\end{equation}
where $F_\ga=\frac{\pa F}{\pa\ck\ga}, F_\be=\frac{\pa
F}{\pa\ck\be}$. In the Hamiltonian form we have
\begin{equation}
\hbar i \pa_t\phi=\hat H\phi,  \quad \hat H\equiv
\vec\al\cdot(-\hbar i \nb -e\vec A)+eA_0+(\mu c
-F_\ga)\ga-F_\be\be.\label{1.6}
\end{equation}
In this paper we denote $\vec A=(A^1,A^2,A^3)$ to be the spatial
part of a contravariant vector $A^\mu$.

It is easy to check that the current conservation law holds $
\pa_{\mu}q^{\mu}=0$, so we can take the normalizing condition as
follows
\begin{equation}
 \int_{R^3}|\phi|^2d^3x=1.
\label{1.7}\end{equation} The $4-d$ potential produced by spinor
$\phi$ itself takes the following form
\begin{equation}
\pa^\al\pa_\al A^{\mu}=
e\ck\al^\mu=e\phi^+\al^\mu\phi.\label{1.8}\end{equation}

\section{Simplification of the Equation}
\setcounter{equation}{0}

Consider a spinor keeping motionless in an external magnetic field
$\vec B=(0,0,B)$. since the scale of the elementary particle is
very small, we take external field $B$ as a constant. By
$\nb\times \vec A_{ext}=\vec B$, we have the general solution for
external vector potential
\begin{equation} \vec A_{ext}=\frac 1 2
B (-y,x,0)+\nb \Phi,\label{poten} \end{equation} where $\Phi(\vec
x)$ is any given smooth function.

In the spherical coordinate system $(r, \th, \vf)$, we have
\begin{equation}
\vec \sg\cdot \nb= \sg_r\pa_r +
(\sg_\th\pa_\th+\sg_\vf\pa_\vf),\label{2.3}
\end{equation}
where $(\sg_r, \sg_\th, \sg_\vf)$ is given by
\begin{equation}
\left\{\left(\begin{matrix}\cos\th & \sin \th e^{-\vf i}\cr\sin\th
e^{\vf i}
 & -\cos\th\end{matrix}\right),
\frac 1 r \left(\begin{matrix}-\sin\th & \cos\th e^{-\vf
i}\cr\cos\th e^{\vf i}
 & \sin\th\end{matrix}\right),\frac 1 {r\sin\th} \left(\begin{matrix}0 &-i e^{-\vf i}\cr i
e^{\vf i} & 0\end{matrix}\right)\right\}.\label{2.4}
\end{equation}
Let $\hat J$ be the angular momentum operator for the spinor field
\begin{equation}
\hat J=\vec r \times (-\hbar i\nb)+\frac 1 2 \hbar \vec S,\qquad
\vec S=\diag (\vec\sg,\vec\sg), \label{2.1}
\end{equation}
then any eigenfunction of $\hat J_3=-\hbar i \pa_\vf+\frac 1 2
\hbar S_3$ takes the following form
\begin{equation}\phi=(u_1,u_2e^{\vf i},-iv_1,-iv_2e^{\vf i})^T \exp\left(\kp \vf
i-\frac{mc^2} \hbar ti\right)\label{2.2}\end{equation} with
$(\kp=0,\pm 1,\pm 2,\cdots)$, where $u_k, v_k(k=1,2)$ are
functions of $r, \th$ but independent on $\vf$ and $t$. In this
paper the index $T$ stands for transposed matrix.

For any spin-$\frac 1 2$ particle, it has a pole axis. If we set
the pole axis as coordinate $x^3=z$, then $\hat J_3$ is
commutative with the nonlinear Hamilton operator (\ref{1.6}) by a
$U(1)$ gauge transformation for spinor as $e^{\Phi i} \phi$, which
removes the uncertain function from external vector potential
(\ref{poten}), thereby we have
\begin{equation} \vec A_{ext}=\frac 1 2
B (-y,x,0)=\frac 1 2 Br\sin\th (-\sin\vf,\cos\vf,0).\label{2.5}
\end{equation}
For the above symmetric form of vector potential (\ref{2.5}),
substituting (\ref{2.2}) into (\ref{1.6}) we can check that all
functions $u_k, v_k$ can take real number. For the vector
potential produced by $\phi$ itself, we will find below it also
takes the form of (\ref{2.5}). This simplification of $\phi$ may
be the essence of the gauge symmetry.

In what follows, we set $\hbar=c=1$ as units for convenience.
Making variable transformation
\begin{equation}
u=u_1(r,\th)+u_2(r,\th)i,\qquad
v=v_1(r,\th)+v_2(r,\th)i,\label{2.6}
\end{equation}
then we have
\begin{equation}\left\{\begin{array}{ll}
\ck\al_0=|u|^2+|v|^2,&\quad \ck\al=(\bar u v-u\bar v)i,\\
\ck\ga~=|u|^2-|v|^2,&\quad \ck\be=\bar u v+u\bar v,
\end{array}\right. \label{2.7}
\end{equation}
with $\ck{\vec\al}=\ck\al(-\sin\vf,\cos\vf,0)$. Substituting
(\ref{2.6}) and (\ref{2.7}) into (\ref{1.3}) we get the Lagrangian
of the eigen states as follows
\begin{eqnarray}
\begin{array}{lll}
{\cal L}& =& \mbox{Re}\left[e^{\th i}\left( \bar u(\pa_r+\frac i r
\pa_\th)\bar v-\bar v (\pa_r+\frac i r \pa_\th)\bar
u\right)\right]
-\frac i {r\sin\th}(\kp+\frac 1 2 )(\bar u v-u\bar v)\\
&&+(m-eA_0)(|u|^2+|v|^2)-ie(\bar u v-u\bar v)A-\mu(|u|^2-|v|^2)
+F,
\end{array} \label{2.8}\end{eqnarray}
where
\begin{equation}
A=\frac 1 r (\vec r \times \vec A)\cdot {\bf e}_z=(\cos\vf
A_y-\sin\vf A_x)=A(r,\sin\th)\label{2.9}\end{equation} including
both external and inner vector potential. By variation with
respect to $\bar u, \bar v$, we get an elegant equation with
double-helix structure
\begin{eqnarray}\left\{
\begin{array}{l}
e^{\th i} (\pa_r+\frac i r \pa_\th)\bar u =\frac i {r\sin\th
}[(\kp+\frac 1 2 )u-\frac 1 2 \bar u]+(\mu+m-eA_0-F_\ga)v+F_\be u
+ie A u,\\
e^{\th i} (\pa_r+\frac i r \pa_\th)\bar v =\frac i {r\sin\th
}[(\kp+\frac 1 2 )v-\frac 1 2 \bar v~]+(\mu-m+eA_0-F_\ga)u-F_\be v
+ie A v.
\end{array} \right.\label{2.10}\end{eqnarray}

By (\ref{2.10}) we find that, $\kp=0$ corresponds to spin $\frac 1
2$ and $\kp=-1$ corresponds to spin $-\frac 1 2$. Generally the
coordinates $r$ and $\th$ can not be separable for nonlinear
equation. The energy functional for (\ref{2.10}) is given by
\begin{eqnarray}\begin{array}{lll}E&=&2\pi\int r^2\sin\th dr
d\th\left\{-\mbox{Re}\left[e^{\th i}\left( \bar u(\pa_r+\frac i r
\pa_\th)\bar v-\bar v (\pa_r+\frac i r
\pa_\th)\bar u\right) \right]-F\right.\\
&&\left.+\left[\frac 1 {r\sin\th}(\kp+\frac 1 2)+eA\right]i(\bar u
v-u\bar
v)+eA_0(|u|^2+|v|^2)+\mu(|u|^2-|v|^2)\right\}.\label{2.11}\end{array}\end{eqnarray}
The eigen solution of (\ref{2.10}) is just the extreme point of
$E$ under the constraint of normalizing condition
\begin{equation} 2\pi\int (|u|^2+|v|^2)r^2\sin\th dr
d\th=1. \label{2.12}\end{equation} (\ref{2.12}) is also the
quantizing condition of the energy spectrum\cite{gu1,gu2}.

\section{A Scheme for Solving Solution}
\setcounter{equation}{0}

For general potential $A^\mu$ and $F$, the analytic solution of
(\ref{2.10}) $u$ and $v$ can not be solved. However they can be
conveniently expressed by Fourier series of $\th$, and the
equations of the radial functions can be derived by variation
principle. For any given integer $N\ge 0$, define $2N+1$ vectors
\begin{eqnarray}
 \Ga(\th) &=& (e^{-2N\th i}, e^{-2(N-1)\th i}, \cdots, e^{2(N-1)\th i},
e^{2N\th i}),\label{3.1}\\
 U(r) &=& (U_{-N}(r), U_{-(N-1)}(r), \cdots, U_{(N-1)}(r), U_{N}(r))^T,\label{3.2}\\
 V(r) &=& (V_{-N}(r), V_{-(N-1)}(r), \cdots, V_{(N-1)}(r),
 V_{N}(r))^T.\label{3.3}
\end{eqnarray}
The eigen solution of (\ref{2.10}) with even parity must take the
form
\begin{equation} u=\Ga\cdot U=\sum^N_{n=-N}U_{n}e^{2n\th i},\quad v=\bar\Ga\cdot V e^{\th i}=\sum^N_{n=-N}V_{n}e^{(-2n+1)\th
i},\label{3.4}
\end{equation}
and the eigen solution with odd parity takes
\begin{equation} u=\Ga\cdot U e^{\th i}=\sum^N_{n=-N}U_{n}e^{(2n+1)\th i},
\quad v=\bar\Ga\cdot V=\sum^N_{-N}V_{n}e^{-2n\th i}.\label{3.5}
\end{equation}
In what follows we only consider (\ref{3.4}). For (\ref{3.5}) we
have similar results.

For the cases $\kp\neq 0$ and $\kp\neq -1$, i.e. for the cases
with nonzero magnetic quantum number, the solution must have
consistent conditions at $\th=0, \pi$ as
\begin{equation} u(r,0)=u(r,\pi)=\sum^N_{n=-N}U_{n}(r)\equiv 0,\quad v(r,0)=v(r,\pi)=\sum^N_{n=-N}V_{n}(r)\equiv0. \label{3.6}
\end{equation}
For this case, (\ref{3.4}) minus (\ref{3.6}) we get
\begin{equation} u=\sum^N_{n=-N}U_{n}(e^{2n\th i}-1),\quad v=\sum^N_{n=-N}V_{n}(e^{-2n\th i}-1)e^{\th
i}.\label{3.7}
\end{equation}
By the form (\ref{3.7}) we have
\begin{equation}
\frac {e^{2n\th i}-1}{2i\sin\th}=(1+e^{2\th i}+e^{4\th
i}+\cdots+e^{2(n-1)\th i})e^{\th i},\label{eee}
\end{equation}
which removes the singularity of (\ref{2.10}) at $\th=0, \pi$.

For the spin $\frac 1 2$ state, i.e. for the case $\kp=0$, we can
check from (\ref{2.10}) that the solution $U, V$ are all real
functions. But for the spin $-\frac 1 2$ state, i.e. for the case
$\kp=-1$, the solution $U, V$ are all pure imaginary functions. In
what follows we only consider the real case. For the covariant
quadratic forms (\ref{2.7}), by (\ref{3.3}) we have
\begin{equation}\left\{\begin{array}{ll}
\ck\al_0=U^T P U+V^T\bar P V,&\quad \ck\al=U^T(Q^+-Q)V i,\\
\ck\ga~=U^TP U-V^T\bar P V,&\quad \ck\be=U^T (Q^++Q)V,
\end{array}\right. \label{3.8}
\end{equation}
where $P=\Ga^+\Ga$ and $Q=\Ga^T\Ga e^{\th i}$ are
$(2N+1)\times(2N+1)$ matrices with components as
\begin{equation}\left\{ \begin{array}{l} P_{m,n}=\exp[2(n-m)\th i],\quad (-N\le n, m \le
N),\\
Q_{m,n}=\exp[(2(n+m)+1)\th i].\end{array}\right.\label{3.9}
\end{equation}
By (\ref{3.8}) and (\ref{3.9}) we have
\begin{eqnarray}
\ck\al_0&=&\sum^N_{n,m=-N}\frac 1 2 (U_n U_m+V_n V_m)(e^{-2(n-m)\th i}+e^{2(n-m)\th i}),\label{3.10}\\
\ck\ga~&=&\sum^N_{n,m=-N}\frac 1 2 (U_n U_m-V_n V_m)(e^{-2(n-m)\th i}+e^{2(n-m)\th i}),\label{3.11}\\
\ck\al &=&\sum^N_{n,m=-N} U_n V_m(e^{-2(n-m)\th i}-e^{2(n-m-1)\th i})e^{\th i}i,\label{3.12}\\
\ck\be &=&\sum^N_{n,m=-N} U_n V_m(e^{-2(n-m)\th i}+e^{2(n-m-1)\th
i})e^{\th i}.
\end{eqnarray} \label{3.13}
The dynamic equation of the potential $A$ is given by
\begin{eqnarray}-\Dl A=e
\ck\al&=&e[a_0(r)\sin\th+a_1(r)\sin3\th+\cdots],\nn\\
&=&e[a_0(r)(e^{-2\th i}-1)+a_1(r)(e^{-4\th i}-e^{2\th
i})+\cdots]e^{\th i}i.\label{3.14}\end{eqnarray}

Substituting (\ref{3.4}), (\ref{3.7}), (\ref{eee}) and
(\ref{3.10})$\sim$(\ref{3.14}) into (\ref{2.10}) and directly
comparing the coefficients of all $e^{2n\th i}$, we can easily get
a truncated ordinary differential equation of $U(r), V(r)$.
However the convergence this cut-off equation to the original
solution needs proof. A more credible method to get the efficient
equation of $U(r), V(r)$ is via variational principle. The
variational equation can be obtained by the following procedure.
Define operators
\begin{equation}\widehat{T}_u=\frac 1 2 \int^\pi_0\Ga^T(\th) e^{-\th
i}\sin\th d\th,\quad \widehat{T}_v=\frac 1 2 \int^\pi_0
\Ga^+(\th)\sin\th d\th, \label{3.15}\end{equation} then
$\widehat{T}_u$ left multiplying the first equation of
(\ref{2.10}) and $\widehat{T}_v$ left multiplying the second give
the variational equation of $U(r), V(r)$. The coefficient matrix
of $U'(r)$ and $V'(r)$ is the same positive definite symmetric
matrix with components
\begin{equation}
M_{n,m}=\frac 1 2 \int^\pi_0 P_{n,m}\sin\th d\th= \frac 1
{1-4(n-m)^2}.\label{3.16} \end{equation} The other coefficient
matrices can also be similarly obtained.

The convergence of expansion (\ref{3.4}) and the consistent
condition (\ref{3.6}) seem to have closely relation with the
structure of nonlinear potential $F(\ck\ga, \ck\be)$, which
reflects the important properties of the elementary particles such
as the exclusion principle. The above procedure is valid for
extensive models. The neat and elegant results are profoundly
rooted in the quaternionic structure of the physical variables and
spacetime\cite{gu9,gu10,21}, so the $3+1$ dimensional Universe is
a miraculous masterpiece with unique feature.


\begin{thebibliography}{99}
\bibitem{1} D. Iuanenko, Sov. Phys. {\bf 13}, 141-149 (1938)
\bibitem{2} H. Weyl, Phys. Rev. {\bf 77}, 699-701 (1950)
\bibitem{3} W. Heisenberg, Physica {\bf 19}, 897-908 (1953)
\bibitem{4} K. Johnson, Phys. Lett. {\bf 78B}, 259-262 (1978)
\bibitem{5} P. Mathieu, Phys. Rev. {\bf D29}, 2879-2883 (1984)
\bibitem{6} A. F. Ranada, {\em Classical nonlinear Dirac field models of extended particles},
In:{\em Quantum theory, groups, fields and particles}, edited by
A.O.Barut, Amsterdam, Reidel, 1983
\bibitem{7} R. Finkelsten, et al, Phys. Rev. {\bf 83(2)}, 326-332(1951)
\bibitem{8} M. Soler, Phys. Rev. {\bf D1(10)}, 2766-2767(1970)
\bibitem{gu1} Y. Q. Gu, {\em Some Properties of the Spinor
Soliton}, Adv in Appl. Clif. Alg. {\bf 8(1)}, 17-29(1998),
http://www.clifford-algebras.org/v8/81/gu81.pdf
\bibitem{gu2} Y. Q. Gu, {\em Characteristic Functions and Typical Values of the Nonlinear Dark
Spinor}, arXiv:hep-th/0611210
\bibitem{gu3} Y. Q. Gu, {\em A Cosmological Model with Dark Spinor Source}, arXiv:gr-qc/0610147
\bibitem{gu4} Y. Q. Gu, {\em Accelerating Expansion of the Universe with Nonlinear Spinors},\\ arXiv:gr-qc/0612176
\bibitem{12} T. Cazenave, L. Vazquez, Comm. Math. Phys.{\bf 105}, 35-47 (1986)
\bibitem{13} F. Merle, J. Diff. Eq. {\bf 74}, 50-68 (1988)
\bibitem{14} M. Balabane, et al, Comm. Math. Phys. {\bf 119}, 153-176 (1988)
\bibitem{15} M. Balabane, et al, Comm. Math. Phys. {\bf 133}, 53-74 (1990)
\bibitem{16} M. J. Esteban, E. Sere, C. R. Acad. Sci. Pavis, {\bf t. 319}, Serie I, 1213-1218 (1994)
\bibitem{17} M. J. Esteban, E. Sere, Comm. Math. Phys. {\bf 171}, 323-350(1995)
\bibitem{18} Wilhelm Fushchych, Renat Zhdanov, {\em SYMMETRIES AND EXACT SOLUTIONS OF NONLINEAR DIRAC
EQUATIONS}, Kyiv Mathematical Ukraina Publisher, Ukraine (1997),
arXiv:math-ph/0609052
\bibitem {19} A. Garrett Lisi, {\em A Solution of the Maxwell-Dirac Equations in
3+1 Dimensions}, arXiv:hep-th/9410244
\bibitem {20} M. Wakano, Prog. Theor. Phys. {\bf 35(6)}, 1117-1141(1996)
\bibitem{gu5} Y. Q. Gu, {\em Spinor Soliton with Electromagnetic Field},
Adv in Appl. Clif. Alg. {\bf V8(2)}, 271-282(1998),
http://www.clifford-algebras.org/v8/82/gu82.pdf
\bibitem{bar1} A. O. Barut and J. Kraus, Found. Physics {\bf 13}, 189 (1983)
\bibitem{bar2} A. O. Barut and J. F. Van Huele, Phys. Rev. {\bf A 32}, 3187(1985)
\bibitem{gu6} Y. Q. Gu, {\em The Electromagnetic Potential Among Nonrelativistic
Electrons}, Adv in Appl. Clif. Alg. {\bf V9(1)}, 55-60(1999),\\
http://www.clifford-algebras.org/v9/91/gu91.pdf
\bibitem{gu7} Y. Q. Gu, {\em New Approach to N-body Relativistic Quantum Mechanics},\\ arXiv:hep-th/0610153
\bibitem{gu8} Y. Q. Gu, {\em Mass Energy Relation of the Nonlinear Spinor}, arXiv:hep-th/0701030
\bibitem{gu9} Y. Q. Gu, {\em A Canonical Form For Relativistic Dynamic
Equation}, Adv in Appl. Clif. Alg. {\bf V7(1)}, 13-24(1997),
\\ http://www.clifford-algebras.org/v7/v71/GU71.pdf, arXiv:hep-th/0610189
\bibitem{gu10} Y. Q. Gu, {\em Green Functions of Relativistic Field Equations}, arXiv:hep-th/0612214
\bibitem{21} A. Gsponer, J.-P. Hurni, {\em quaternions in mathematical physics \\ (1):
Alphabetical bibliography}, arXiv:math-ph/0510059; \\ {\em (2):
Analytical bibliography},  arXiv:math-ph/0511092.
\end{thebibliography}
\end{document}